\documentclass[doublespace,fleqn,10pt]{wlscirep}

\usepackage[utf8]{inputenc}
\usepackage[T1]{fontenc}
\usepackage{graphicx}
\usepackage{mathtools}
\usepackage{xcolor}
\usepackage{amsmath}
\usepackage{amssymb}
\usepackage{xfrac}
\usepackage{soul}
\usepackage{mathrsfs}
\usepackage{bm}
\usepackage{siunitx}
\usepackage{booktabs}
\usepackage{multirow}
\usepackage{makecell}
\usepackage{subfigure}
\usepackage[section]{placeins}
\hypersetup{colorlinks=true,linkcolor=blue,citecolor=blue,urlcolor=blue}

\begin{document}

\title{Interface-Controlled Phase Stability in Polymorphic HfO$_2$ Revealed by Machine-Learning Atomistic Simulations}

\author[1,*]{Xudong Zhu}
\author[2,3]{Junhong Li}
\author[1,2,3,*]{Lixin He}

\affil[1]{Institute of Artificial Intelligence, Hefei Comprehensive National Science Center, Hefei, Anhui, 230088, People's Republic of China}
\affil[2]{Laboratory of Quantum Information, University of Science and Technology of China, Hefei, 230026, People's Republic of China}
\affil[3]{Hefei National Laboratory, University of Science and Technology of China, Hefei, Anhui, 230088, People's Republic of China}

\affil[*]{Correspondence: zhu@iai.ustc.edu.cn; helx@ustc.edu.cn}

\begin{abstract}

HfO$_2$ exhibits rich polymorphism, and competition among different phases underpins many of its functional properties. Yet bulk free-energy relations alone cannot explain phase selection at mixed-phase boundaries, where interface orientation and structural continuity constrain collective rearrangements. Here, using machine-learning atomistic simulations and a Hf-centered local phase classification scheme, we show that crystallographic interface matching redirects phase competition and accessible transformation pathways. The M(100)/T(100) interface remains pinned as an M/T mixture throughout 3~ns simulations from 300 to 1800~K. M/PO, M/AO, and PO/AO interfaces retain two-phase coexistence up to 900~K, whereas all T/PO interfaces become PO-dominant. At 1800~K, all non-M interfaces become T-dominant, while M-containing interfaces retain a monoclinic majority.
Nudged elastic band calculations reveal lower-barrier routes through interface states. For M(100)$\rightarrow$T(100), the interface-mediated route in a long cell ($\sim$12~nm) yields a barrier of 136.03~meV/f.u., 29.2\% lower than the direct route in a short cell ($\sim$3~nm). This difference is associated with sequential phase-front motion absent from the short cell. These results identify phase boundaries as active participants in phase stability and transformation and establish interface orientation and crystallographic matching as variables for stabilizing metastable polymorphs and directing phase conversion in HfO$_2$.

\end{abstract}

\date{\today}
\maketitle

\section*{Introduction}
Hafnia (HfO$_2$) and its solid solutions are central oxides in advanced complementary metal oxide semiconductor technologies because they combine a high dielectric constant, ferroelectricity, wide band gap, and compatibility with silicon processing \cite{Robertson2004,Cheema2020,Jung2022,Cheema2022,Song2026}. These properties arise from an unusually rich polymorphism. Alongside the monoclinic (M) ground state, tetragonal (T), cubic (C), polar orthorhombic (PO), antipolar orthorhombic (AO), and other metastable structures can be selected by temperature, stress, composition, defects, surfaces, and electrical boundary conditions \cite{Boescke2011,Mueller2012,Polakowski2015,Park2015,Schroeder2022,Yang2023,Huan2014,ReyesLillo2014,Materlik2015,Batra2016,Raeliarijaona2023,Wang2023,Kumar2024,Shen2024}. Because these phases differ in symmetry, polarization, and dielectric response but remain close in energy, constraints acting over only a few atomic layers can alter the structure retained by a film or grain. Resolving phase competition is therefore essential both for stabilizing useful functional states and for understanding the structural reliability of materials in which several polymorphs coexist.

Bulk calculations and experiments have established how strain, defects, pressure, surfaces, electrodes, and electrostatic conditions shift the relative stability of the major phases \cite{Huan2014,ReyesLillo2014,Materlik2015,Batra2016,Raeliarijaona2023,Wang2023,Kumar2024,Shen2024,Li2025,Xu2017,Starschich2017,Kuenneth2017,Yao2019,Glinchuk2020,Athle2021,He2021,Han2023,Bai2023,Huang2024,Kim2024,Shi2024,Zhao2024,Liu2025,Zhou2025}. Studies of surfaces, grain boundaries, coherent interfaces, and domain walls further demonstrate that a finite region cannot be described by bulk phase energy alone \cite{Materlik2015,Park2017Surface,Xu2024Strain,Ding2020DomainWall,Falkowski2021Interphase,Kelley2023}. Transformation studies add a kinetic dimension: access to T, PO, AO, and M depends on oxygen motion, lattice shear, and collective structural rearrangement \cite{Xu2021,Ma2023,Zhu2024,Zhou2024,Hu2025,Zhou2025b,Zhang2025,Li2025a}. The strong increase in the T$\rightarrow$M barrier when the monoclinic shear is included illustrates how directly the chosen structural constraint can alter a pathway \cite{Xu2021}. Machine-learning potentials now extend this analysis from small cells to amorphous and liquid hafnia, bulk thermodynamics, phase transformations, and field-driven dynamics \cite{Sivaraman2020,Wu2021,Bichelmaier2023,Bichelmaier2024,Ouyang2024,Lee2026FieldInduced}. Together, these advances establish the thermodynamic, interfacial, and kinetic factors governing phase selection, yet how a specific interface orientation directs this process at finite temperatures remains to be determined.

A coherent boundary between two hafnia polymorphs introduces a structural problem that bulk phase diagrams and small transformation cells cannot resolve. The outcome depends not only on the identities of the two parent phases, but also on crystal orientation, coherent strain, interface termination, and the distance available for structural accommodation. These constraints can preserve both parent phases, favor one phase, or create an extended arrangement that is absent from the initial two-phase label \cite{Alcala2023,Shi2023,Kelley2023,Mukherjee2024}. Such behavior is especially relevant because mixed-phase regions are common in experimental hafnia \cite{Wang2023,Huang2024,Liu2025,Zhou2025}. Small periodic cells restrict long-range relaxation and the motion of an extended phase front \cite{Mu2025,Geng2026}, whereas direct \textit{ab initio} molecular dynamics (AIMD) cannot combine nanometer lattice with nanosecond simulation. The remaining question is therefore how boundary geometry, bulk thermodynamic bias, and pathway accessibility work together to select and retain local phases in a realistic mixed-phase region.

In this work, we investigate interface-controlled phase selection in HfO$_2$ using machine-learning atomistic simulations and a Hf-centered local phase classification scheme. We train a Deep Potential (DP) force field and validate its accuracy against both bulk and interfacial density functional theory (DFT) data. Finite-temperature simulations are performed for 45 coherent two-phase interfaces constructed from M, T, PO, and AO crystals with different crystallographic orientations. We further develop a 21-component local descriptor to resolve the spatial distribution of the four phases in HfO$_2$, enabling the identification of coexisting polymorphs. Starting from the bulk thermodynamic baseline, we examine how interface orientation modifies local phase selection from 300 to 1800~K and whether the resulting structures persist on nanosecond time scales. The M(100)/T(100) interface provides a representative phenomenon of pinning: a mixed M/T state persists for 3~ns, preserving T against the low-temperature preference for M while retaining M as the bulk free-energy preference shifts toward T upon heating. Across the full interface set, crystallographic matching determines the coexistence of parent phases. We then use nudged elastic band calculations to investigate transformation pathways connecting interface states to their neighboring pure phases. This combined analysis distinguishes temperature-dependent bulk free-energy preferences from the structural accessibility imposed by a specific interface and shows that sufficiently extended cells permit sequential phase-front motion that is inaccessible to smaller cooperative cells. Our work links crystallographic matching, local phase persistence, and transformation-pathway accessibility, providing an atomistic basis for controlling metastable HfO$_2$ polymorphs through interface design.

\section*{Results}

\subsection*{Bulk Thermodynamic Baseline}

The bulk thermodynamic baseline was obtained from quasi-harmonic approximation (QHA) calculations for the M, T, PO, AO, and C phases using DFT and the DP force field, as described in Methods.

\begin{figure}[htbp]
	\centering
	\includegraphics[width=0.48\textwidth]{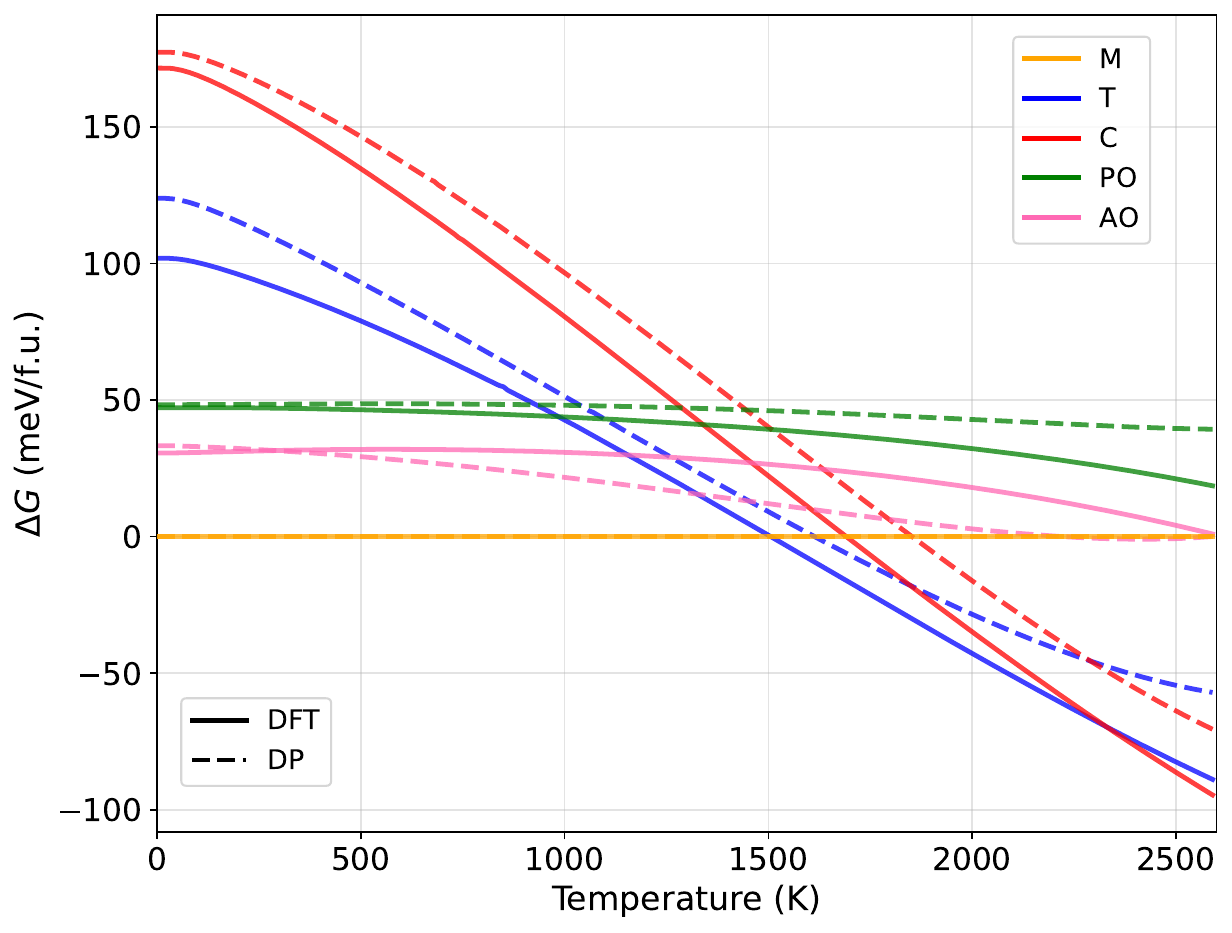}
	\caption{\textbf{Relative Gibbs free energies of bulk HfO$_2$ polymorphs.} Relative Gibbs free energies, $\Delta G_{\alpha}(T)=G_{\alpha}(T)-G_{\mathrm{M}}(T)$, of the M, T, C, PO, and AO phases calculated within the quasi-harmonic approximation using DFT and the DP force field. Solid and dashed curves denote the DFT and DP force field results, respectively. Orange, blue, red, green, and pink denote the M, T, C, PO, and AO phases, respectively. The M phase defines the zero-energy reference at each temperature. Energies are reported in meV per formula unit (f.u.).}
	\label{fig:bulk_gibbs}
\end{figure}

The resulting relative Gibbs free energies are shown in Fig.~\ref{fig:bulk_gibbs}. DFT and the DP force field yield a consistent bulk thermodynamic landscape. At low temperature, M exhibits the lowest free energy, while PO and AO remain metastable and closely spaced in energy. As temperature increases, the relative free energies of T and C decrease more rapidly due to their vibrational contributions, progressively enhancing their thermodynamic stability relative to the low-temperature phases. The proximity of the PO and AO free energies indicates that even modest local structural constraints may alter their relative accessibility without changing the overall bulk phase hierarchy.

The characteristic free-energy crossovers provide a more stringent assessment of the DP force field than does the phase ordering at a single temperature. In the DFT results, T becomes lower in free energy than PO at approximately 1000~K and lower than AO at approximately 1200~K. M remains the lowest-free-energy bulk phase up to about 1500~K, beyond which T becomes thermodynamically favored, while C becomes competitive only at still higher temperatures. The DP force field reproduces the same sequence with moderate shifts in the crossover temperatures. It therefore captures not only the low-temperature stability of M but also the progressive stabilization of T and C and the relative proximity of the two orthorhombic phases. This calculated sequence agrees with the experimental progression of pure HfO$_2$ from the monoclinic phase toward tetragonal and cubic structures at high temperature, for which the reported M$\rightarrow$T and T$\rightarrow$C transition temperatures are approximately 1973~K and 2803~K, respectively \cite{Hong2018}. The calculated crossover temperatures should not be interpreted as exact experimental transition temperatures, because the QHA describes ideal bulk crystals and incorporates anharmonicity only through volume-dependent phonons. Nevertheless, these relations establish the bulk free-energy preference relevant to the interface simulations: heating progressively favors T over PO and AO, and eventually over M.

The bulk free-energy relations quantify the intrinsic thermodynamic preferences among ideal HfO$_2$ polymorphs but do not include boundary-specific contributions arising from crystallographic orientation, strain, and interfacial reconstruction. They therefore define the thermodynamic preference of the bulk phases but are insufficient to determine phase selection in coherently constrained two-phase structures.

\subsection*{Interface-Dependent Phase Stability at 300 K}

To determine how crystallographic interface matching modifies phase-boundary stability at 300~K, we analyze the local phase distributions of 45 coherent two-phase models constructed from M, T, PO, and AO phases, comprising six phase-pair families and the orientations defined in Methods. Local phase fractions and Hf-site-resolved maps were extracted from finite-temperature trajectories using the same four-phase classification scheme. The DP force field accuracy in interfacial configuration space was validated by two complementary DFT benchmarks: evaluation of 12,006 frames from AIMD trajectories yielded an energy MAE of 2.15~meV/atom and a force MAE of 0.17~eV/\AA, while DFT validation of 600 configurations from independent DPMD trajectories gave an energy MAE of 1.05~meV/atom and a force MAE of 0.17~eV/\AA. The comparable errors confirm that the force field reproduces both AIMD-sampled and self-generated interfacial structures, complementing the bulk benchmarks reported in Figs.~S1--S5.

\begin{figure*}[htbp]
	\centering
	\includegraphics[width=1.0\textwidth]{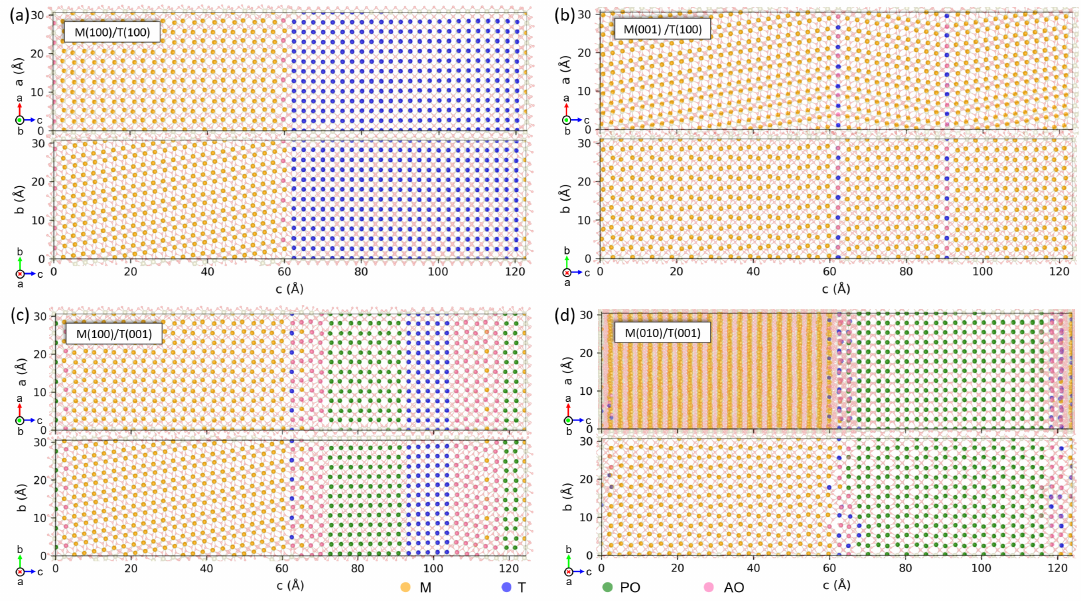}
	\caption{\textbf{Representative M/T interface structures after the 300~K finite-temperature protocol.} For each interface, the upper and lower panels show projections onto the $ac$ and $bc$ planes, respectively. Phase-colored Hf sites are assigned using the M, T, PO, and AO local descriptor classification. (a) M(100)/T(100) retains two extended parent domains separated by a relatively sharp boundary. (b) M(001)/T(100) transforms predominantly into M, leaving a narrow T-like boundary. (c) M(100)/T(001) develops a reconstructed layer containing M, PO, AO, and residual T environments. (d) M(010)/T(001) reconstructs into a mixed stacking dominated by M and PO environments.}
	\label{fig:represent_MT_interface}
\end{figure*}

The M/T family results shown in Fig.~\ref{fig:represent_MT_interface} illustrate the sensitivity of phase selection to crystallographic interface matching. In M(100)/T(100) [Fig.~\ref{fig:represent_MT_interface}(a)], both parent domains remain spatially extended with a narrow boundary, and the nearly balanced M and T fractions indicate stable M/T coexistence at 300~K. By contrast, M(001)/T(100) [Fig.~\ref{fig:represent_MT_interface}(b)] transforms almost completely into M, leaving only a narrow T-like region between monoclinic domains of different orientations. The remaining M/T results go beyond simple parent preservation or conversion: M(100)/T(001) [Fig.~\ref{fig:represent_MT_interface}(c)] develops a broad reconstructed region containing M, PO, AO, and residual T, while M(010)/T(001) [Fig.~\ref{fig:represent_MT_interface}(d)] forms a more ordered stacking dominated by M and PO with little residual T. These reconstructed regions extend beyond a narrow fluctuating boundary and contain arrangements absent from the initial lattices. Thus, different crystallographic matchings within the same phase pair produce three distinct responses: retained M/T coexistence, conversion toward M, and reconstruction into mixed monoclinic/orthorhombic environments.

\begin{figure*}[htbp]
	\centering
	\includegraphics[width=1.0\textwidth]{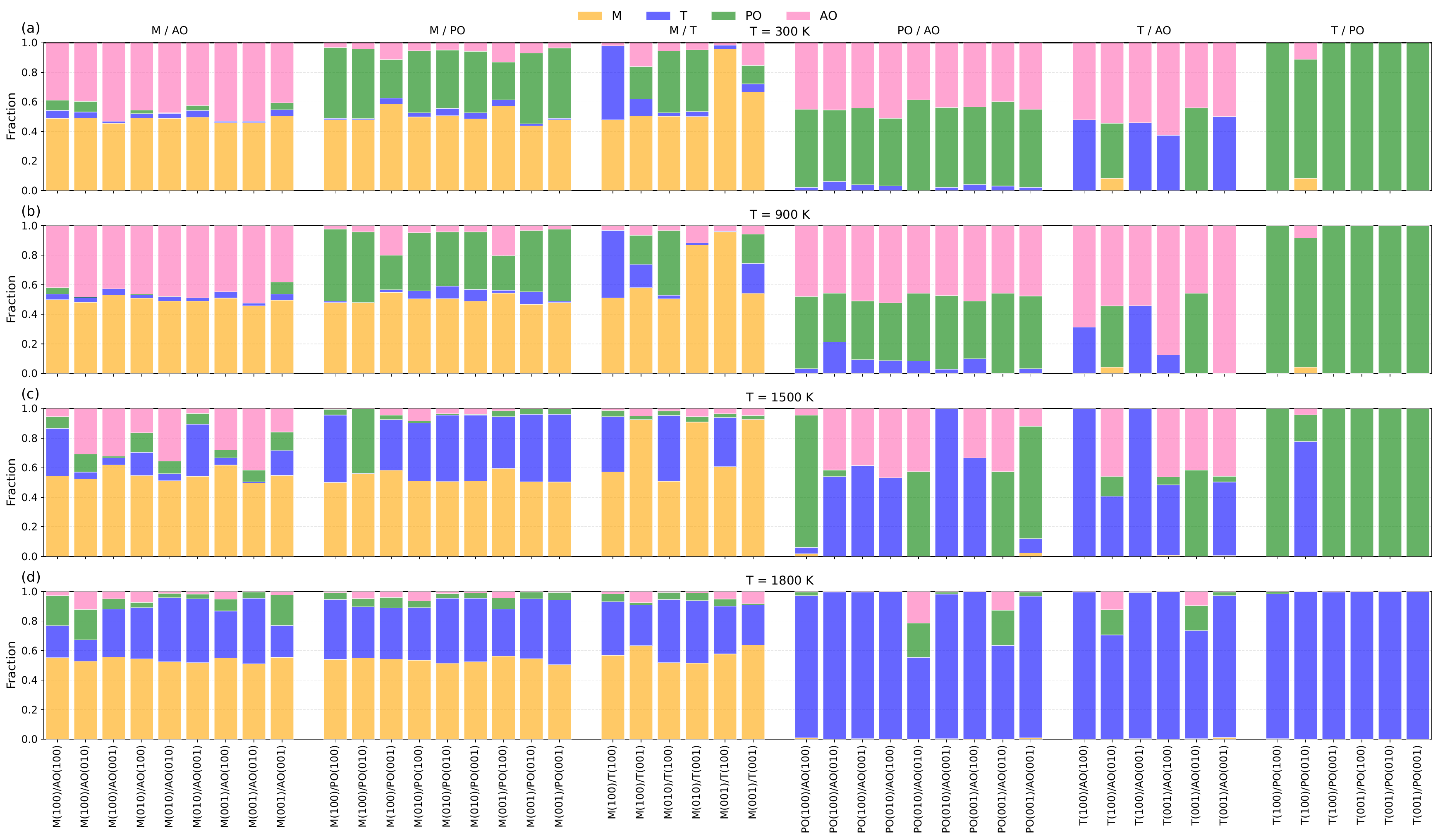}
	\caption{\textbf{Phase fractions of the two-phase interface models at representative temperatures.} Phase fractions of 45 two-phase interface models at (a) 300~K, (b) 900~K, (c) 1500~K, and (d) 1800~K. The models are grouped by initial parent phase pair into the M/AO, M/PO, M/T, PO/AO, T/AO, and T/PO families. Colors indicate the fractions of Hf sites assigned to local M (orange), T (blue), PO (green), and AO (pink) environments by the 21-component local phase classifier.}
	\label{fig:phase_fraction_summary}
\end{figure*}

The phase fraction distributions in Fig.~\ref{fig:phase_fraction_summary}(a) show how the low-temperature thermodynamic preference for M is filtered by the coherent constraints of each crystallographic matching. All nine M/AO interfaces retain substantial fractions of both parent phases, with M fractions of 0.455--0.504 and AO fractions of 0.389--0.532. All M/PO interfaces likewise preserve both M and PO. Seven remain close to balanced mixtures, whereas M(001)/PO(100) and M(100)/PO(001) become more M-rich, with M fractions of 0.572 and 0.586, respectively, and also develop additional AO character. The emergence of AO suggests that part of the initially parallel oxygen displacement order in PO becomes locally compensated or reversed under the coherent constraint, rather than undergoing a simple PO$\rightarrow$M conversion. The M/T family shows a much stronger dependence on crystallographic interface matching. M(100)/T(100) retains an almost equal mixture, with M and T fractions of 0.479 and 0.500, whereas M(001)/T(100) transforms almost completely into M, reaching an M fraction of 0.958. The other four M/T interfaces lose most of their initial T character and form either M-rich or mixed M/orthorhombic configurations, with residual T fractions of only 0.024--0.115. Thus, the bulk free energy of M favors monoclinic reconstruction, but the interface matching determines whether the associated transformation strain and oxygen sublattice rearrangement propagate through the T region, become pinned at the boundary, or are redirected toward an orthorhombic configuration.

The non-M families reveal a complementary role of polar oxygen ordering. All T/PO interfaces lose their initial T character: five become entirely PO, while T(100)/PO(010) shows a PO fraction of 0.805 and smaller AO and M fractions of 0.112 and 0.083. This systematic preference for PO suggests that its ordered polar oxygen displacements provide a structural template from which polar order propagates into the adjoining nonpolar T region. The T/AO family behaves differently because the antiparallel oxygen displacements of bulk AO cancel macroscopically and provide no equivalent uniform polar template. Four of the six interfaces consequently retain substantial fractions of both parent phases, with T fractions of 0.375--0.500 and AO fractions of 0.500--0.625. The two interfaces containing AO(010), however, lose their T component and reconstruct into PO/AO-rich configurations. In the AO(010) region, the two periodic boundaries are structurally inequivalent: one exposes a locally polar arrangement of oxygen displacements within the interface plane, whereas the opposite boundary is locally compensated. The observed reconstruction is consistent with the polar termination promoting a PO-like displacement pattern in the adjoining T region, allowing polar order to propagate through the initially tetragonal domain in a manner analogous to the T/PO interfaces. The initial T/AO coexistence is retained when this termination is absent. All nine PO/AO interfaces also preserve both orthorhombic orderings, with PO fractions of 0.458--0.614 and AO fractions of 0.386--0.510, no detectable M, and at most 0.061 T. Because PO and AO share closely related local coordination but differ in the parallel or antiparallel correlation of their oxygen displacements, their boundaries can accommodate changes in polar order without complete reconstruction of the orthorhombic framework. These results identify the orientation and termination of the oxygen displacement pattern as central factors governing whether an interface retains its parent phases or opens a different transformation pathway.

The preferential reconstruction of T/PO toward PO is consistent with the fixed-lattice first-principles pathways reported by Xu \textit{et al.}, where the T$\rightarrow$PO barrier is 45~meV/f.u. versus 70~meV/f.u. for T$\rightarrow$AO \cite{Xu2021}. Although those values describe smaller bulk cells rather than coherent boundaries, their ordering provides a useful kinetic reference for the near-complete loss of T in T/PO and the larger residual T fraction in T/AO. The present interface dataset adds a crystallographic constraint: even for the same parent pair, interface termination and boundary orientation determine whether a nominally accessible rearrangement proceeds within the simulated interval or is redirected toward another local structure.

The 300~K results thus reveal a two-level selection mechanism. The parent-phase combination defines the competing structural motifs and their intrinsic transformation tendencies, while crystallographic matching controls which pathways remain accessible at a particular coherent boundary. Local phase stability is consequently governed by the combined effects of bulk thermodynamic metastability, pathway accessibility, and interface-specific structural compatibility.

\subsection*{Temperature-Driven Interfacial Phase Evolution}

The temperature series reveals that coherent HfO$_2$ interfaces do not follow a common phase-evolution sequence. Fig.~\ref{fig:phase_fraction_summary} shows the phase distributions at 300, 900, 1500, and 1800~K, while the complete six-temperature series is provided in Fig.~S8. Up to 1200~K, most non-M interfaces remain predominantly orthorhombic. At 1500~K, selected PO/AO and T/AO interfaces reconstruct extensively toward T, whereas most T/PO interfaces remain PO-rich. By 1800~K, all non-M interfaces become T-dominant. M-containing interfaces behave differently: M remains the largest component in every structure, although increasing T populations develop in the adjoining regions. Temperature therefore changes the phases accessible to each interface, but the onset and extent of reconstruction remain determined by the parent structures and their crystallographic matching.

At 900~K [Fig.~\ref{fig:phase_fraction_summary}(b)], the interface states established at lower temperature remain largely intact. All six T/PO interfaces contain no detectable T and have a mean PO fraction of 0.979. This persistence indicates that the polar distortion initially supplied by PO has already propagated through most of the former T region. Recovering T would require the correlated off-centering and associated oxygen-sublattice distortion to be suppressed across an extended PO-like domain, which is not observed on the simulated timescale. T/AO instead retains mean AO and T fractions of 0.684 and 0.149, respectively, while PO/AO contains comparable PO and AO populations of 0.444 and 0.483, together with 0.074 T. Unlike the parallel polar order of PO, the antiparallel displacements in AO and the competing displacement patterns at PO/AO boundaries contain locally compensated regions. These regions can accommodate partial recovery of T-like coordination without requiring reversal of a uniformly aligned polar distortion. This distinction is consistent with the larger residual T population in T/AO than in T/PO.

The M-containing families also remain structurally distinct at 900~K. Their mean M fractions are 0.496 for M/AO, 0.500 for M/PO, and 0.661 for M/T, while M/AO and M/PO retain substantial fractions of their orthorhombic parent phases. The persistence of M cannot be attributed solely to oxygen displacement order. The monoclinic structure also contains a shear distortion, asymmetric Hf--O coordination, and coupled rearrangements of the Hf and O sublattices. Within a coherent two-phase cell, these features form a connected low-symmetry framework whose conversion requires simultaneous changes in local coordination and lattice distortion. Consequently, the free-energy crossover between T and PO near this temperature does not produce either immediate or monotonic conversion of the interface structures. The relative bulk free energies identify which phases become energetically competitive, but do not determine whether the required collective rearrangement is compatible with a particular coherent boundary.

A pronounced orientation dependence emerges at 1500~K [Fig.~\ref{fig:phase_fraction_summary}(c)]. Five of the nine PO/AO interfaces become T-dominant, giving a family-averaged T fraction of 0.388. Four of the six T/AO interfaces are also T-dominant, including two that become entirely T-like, and the mean T fraction reaches 0.563. In these families, cancellation or disruption of the pre-existing polar and antipolar displacement correlations can produce locally T-like coordination without requiring a single transformation front to reverse a uniformly ordered domain. Whether this rearrangement propagates through the cell nevertheless depends on the orientation of the orthorhombic distortion relative to the coherent lattice.

T/PO remains the principal exception at 1500~K. Five orientations are still PO-dominant, and the family retains a mean PO fraction of 0.864. Only T(100)/PO(010) becomes T-dominant. The persistence of the other five structures shows that the PO configuration generated at lower temperature constitutes a robust extended state rather than a weakly perturbed T lattice. Its conversion requires the collective removal of polar off-centering together with recovery of the tetragonal coordination pattern. The distinct behavior of T(100)/PO(010) further shows that this process is sensitive to the orientation of the PO distortion and to its compatibility with the common cell. Thus, even within one phase family at a fixed temperature, crystallographic matching determines whether T-like coordination can nucleate and extend across the structure.

All 24 M-containing interfaces remain M-dominant at 1500~K, with mean M fractions of 0.550, 0.530, and 0.741 for M/AO, M/PO, and M/T, respectively. Their secondary phase populations reveal how the adjoining domains respond while the monoclinic framework persists. Eight of the nine M/PO interfaces develop T fractions of 0.343--0.458, whereas M(100)/PO(010) retains a mixed M/PO structure with almost no T. The orthorhombic region can therefore reconstruct toward T without requiring concurrent removal of the M domain. M/T shows an even clearer orientation dependence. The three interfaces containing T(001) reach M fractions of 0.909--0.928 and contain almost no T, indicating that their matching permits the monoclinic distortion to propagate through the original T region. By contrast, the three T(100) interfaces retain T fractions of 0.333--0.444, consistent with pinning of the M/T boundary. The different outcomes arise from the orientation of the monoclinic shear and coordination distortion relative to the coherent interface, rather than from phase composition alone.

At 1800~K [Fig.~\ref{fig:phase_fraction_summary}(d)], the M-containing and non-M interfaces exhibit sharply different responses. All 21 non-M interfaces are T-dominant, with mean T fractions of 0.897 for PO/AO, 0.896 for T/AO, and 0.994 for T/PO. Although residual orthorhombic environments remain in selected PO/AO and T/AO orientations, the extended PO and AO regions have largely transformed into T. In contrast, all 24 M-containing interfaces remain M-dominant. Their mean M fractions are 0.538, 0.536, and 0.575 for M/AO, M/PO, and M/T, respectively, accompanied by T fractions of 0.320, 0.387, and 0.347. Heating therefore converts much of the non-M regions into T while preserving a connected monoclinic component, identifying the M region as a structural anchor that restricts complete transformation of the coherent cell. Because C becomes competitive at high temperature, the 1800~K assignments were further examined using an auxiliary T/C classifier, which confirms that the representative T-rich interfacial regions retain tetragonal rather than locally cubic character (see Fig.~S10 and Tab.~S2).

The complete temperature series consequently resolves two successive features of interfacial phase evolution. Between 1200 and 1500~K, reconstruction toward T is strongly selective and occurs only for crystallographic matchings that accommodate the required changes in coordination, displacement order, and lattice distortion. At 1800~K, all non-M interfaces become T-dominant, whereas M-containing interfaces preserve a monoclinic framework and develop mixed M/T states. The bulk free-energy relations explain why T becomes progressively more competitive on heating, but the observed conversion sequence is controlled by the structural pathways available within each coherent interface. The temperatures reported here therefore characterize reconstruction within the finite simulation protocol and should not be interpreted as equilibrium phase boundaries.

\begin{figure*}[htbp]
	\centering
	\includegraphics[width=1.0\textwidth]{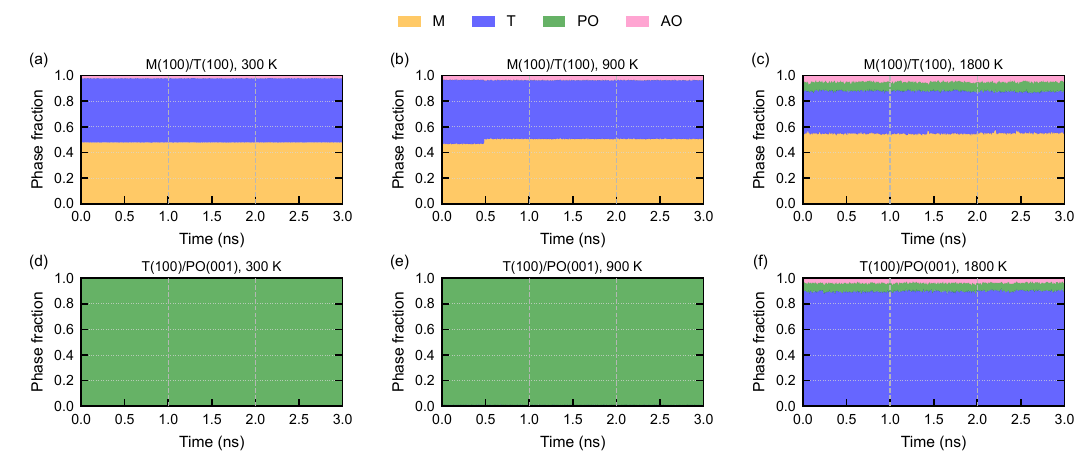}
	\caption{\textbf{Nanosecond phase evolution of two representative interfaces.} (a--c) M(100)/T(100) at (a) 300~K, (b) 900~K, and (c) 1800~K. (d--f) T(100)/PO(001) at (d) 300~K, (e) 900~K, and (f) 1800~K. Stacked areas show the M, T, PO, and AO fractions assigned by the 21-component local phase classifier over nonoverlapping 5~ps windows.}
	\label{fig:phase_fraction_3ns}
\end{figure*}

\subsection*{Nanosecond Persistence of Interface-Controlled States}

To examine the nanosecond stability of representative interface-controlled states, we extended the simulations of M(100)/T(100) and T(100)/PO(001) to 3~ns at 300, 900, and 1800~K. These interfaces represent two complementary outcomes: persistent M/T coexistence and temperature-dependent selection between PO-rich and T-rich states.

M(100)/T(100) maintains substantial fractions of both parent phases at all three temperatures. At 300~K [Fig.~\ref{fig:phase_fraction_3ns}(a)], the M and T fractions fluctuate around mean values of 0.479 and 0.500, respectively, with only 0.021 AO. At 900~K [Fig.~\ref{fig:phase_fraction_3ns}(b)], the balance shifts modestly toward M, yielding mean M and T fractions of 0.500 and 0.464, but neither phase grows continuously at the expense of the other. The interface remains mixed at 1800~K [Fig.~\ref{fig:phase_fraction_3ns}(c)], with mean M, T, PO, and AO fractions of 0.549, 0.330, 0.072, and 0.049, respectively. The final 200~ps contains 0.553 M and 0.324 T, closely matching the full-trajectory averages and showing no late-stage progression toward a homogeneous phase. The reciprocal retention of T when M has the lower bulk free energy at 300~K and of M when T becomes lower in free energy at 1800~K identifies this configuration as a pinned M/T boundary rather than a slowly advancing transformation front. Conversion between M and T requires coupled changes in monoclinic shear, Hf--O coordination, and the oxygen sublattice. For the M(100)/T(100) matching, these rearrangements do not propagate across the boundary within 3~ns. The small PO and AO populations at 1800~K indicate limited local reconstruction without disrupting the extended M/T configuration.

T(100)/PO(001) exhibits a different long-term response. The structure remains almost entirely PO at 300~K [Fig.~\ref{fig:phase_fraction_3ns}(d)] and 900~K [Fig.~\ref{fig:phase_fraction_3ns}(e)], with mean PO fractions of 1.000 and 0.989, respectively, but becomes predominantly T at 1800~K [Fig.~\ref{fig:phase_fraction_3ns}(f)] with a mean T fraction of 0.895. This interface therefore does not preserve long-term coexistence between its parent phases. Instead, it retains an extended PO-rich state at lower temperatures and an extended T-rich state at 1800~K. The low-temperature result is consistent with the PO region templating correlated polar off-centering into the adjoining T region. At 1800~K, this extended displacement order is no longer retained and T-like coordination dominates. Unlike M/T conversion, this reorganization does not require propagation or removal of a connected monoclinic shear distortion, allowing the T/PO structure to select different extended states at different temperatures.

The two interfaces thus reveal complementary forms of long-term phase selection. M(100)/T(100) preserves mixed-phase coexistence across the investigated temperature range, whereas T(100)/PO(001) retains the dominant phase selected within each temperature range. Their behavior agrees with the corresponding family-level results in Fig.~\ref{fig:phase_fraction_summary}, confirming that these interface-controlled states persist over nanoseconds rather than representing transient fluctuations. The long trajectories therefore demonstrate the robustness of interface-controlled phase selection across distinct structural and temperature conditions.

\begin{figure*}[htbp]
	\centering
	\includegraphics[width=0.82\textwidth]{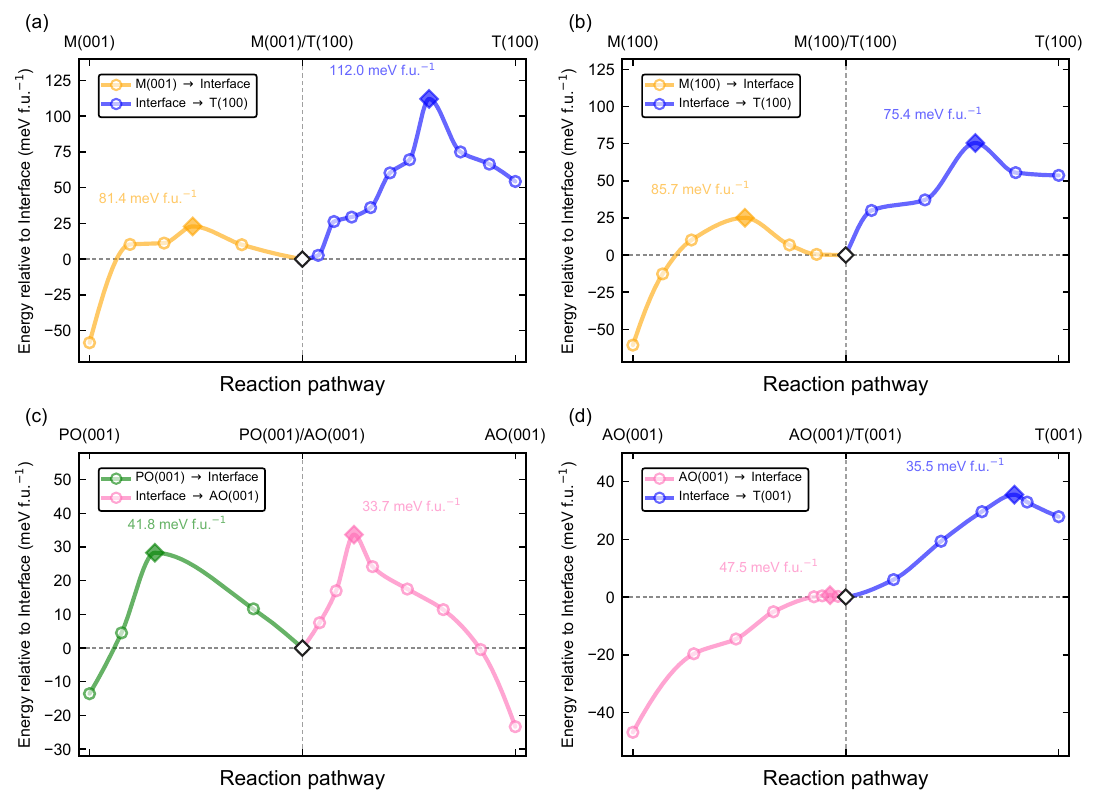}
	\caption{\textbf{Transformation pathways through representative relaxed interface states.} Fixed-lattice energy profiles for (a) M(001)$\rightarrow$T(100), (b) M(100)$\rightarrow$T(100), (c) PO(001)$\rightarrow$AO(001), and (d) AO(001)$\rightarrow$T(001). Energies in each panel are reported relative to the corresponding interface state (I), and the annotations indicate the directional barrier for each pathway segment.}
	\label{fig:interface_escape_neb}
\end{figure*}

\subsection*{Transformation Pathways through Interface States}

The MD simulations establish distinct interface-controlled outcomes, including retained phase coexistence and preferential conversion toward one phase. We next examined the transformation pathways associated with four representative relaxed interface states using fixed-lattice nudged elastic band calculations (Fig.~\ref{fig:interface_escape_neb}). The M(001)/T(100), M(100)/T(100), PO(001)/AO(001), and T(001)/AO(001) interfaces represent conversion toward M, persistent M/T coexistence, PO/AO coexistence, and preferential retention of AO against T, respectively. The relaxed interface state in each structure is labeled I and provides a common energy reference for the two branches connecting it to the corresponding pure-phase endpoints.

The two M/T pathways show a common preference toward M but different directional asymmetries. For M(001)/T(100), the M$\rightarrow$I and I$\rightarrow$T barriers are 81.36 and 112.03~meV per formula unit, respectively, while the reverse I$\rightarrow$M and T$\rightarrow$I barriers are 22.88 and 57.72~meV per formula unit. Reconstruction from the interface state is therefore substantially easier toward M than toward T. M(100)/T(100) shows the same ordering with a smaller difference between the two branches. The M$\rightarrow$I and I$\rightarrow$T barriers are 85.71 and 75.43~meV per formula unit, while the I$\rightarrow$M and T$\rightarrow$I barriers are 25.11 and 21.78~meV per formula unit. The I$\rightarrow$M barriers are similar for the two orientations, whereas the I$\rightarrow$T barrier decreases from 112.03 to 75.43~meV per formula unit. This difference is consistent with their 300~K behavior: M(001)/T(100) converts almost entirely toward M, whereas M(100)/T(100) retains an extended T region and remains mixed over 3~ns. Crystallographic matching therefore primarily modifies the T-directed branch and influences whether the M/T boundary advances or remains pinned.

The orthorhombic pathways show a near-balanced and a strongly asymmetric case. For PO(001)/AO(001), the PO$\rightarrow$I and I$\rightarrow$AO barriers are 41.84 and 33.67~meV per formula unit, while the reverse I$\rightarrow$PO and AO$\rightarrow$I barriers are 28.29 and 57.02~meV per formula unit. The similar I$\rightarrow$PO and I$\rightarrow$AO barriers of 28.29 and 33.67~meV per formula unit indicate no strong preference between the two reconstruction directions, consistent with the retained PO/AO coexistence at lower temperatures. For T(001)/AO(001), the AO$\rightarrow$I and I$\rightarrow$T barriers are 47.49 and 35.46~meV per formula unit, while the reverse T$\rightarrow$I and I$\rightarrow$AO barriers are 7.63 and 0.63~meV per formula unit. Comparison of the two branches leaving the interface state, 0.63~meV per formula unit toward AO and 35.46~meV per formula unit toward T, shows a pronounced preference for reconstruction toward AO. The relaxed interface states are therefore neither common energetic midpoints nor equivalent intermediates across different phase pairs. Their local coordination and displacement patterns determine the barrier associated with each reconstruction direction.

Direct-path controls quantify the reduction associated with the interface-mediated routes. For PO(001)$\rightarrow$AO(001) under the same lattice, the direct pathway reaches 170.35~meV per formula unit above PO, whereas the pathway through the relaxed interface state reaches 47.22~meV per formula unit, a reduction of 72.3\%. For T(001)$\rightarrow$AO(001), the corresponding maxima are 80.70 and 7.62~meV per formula unit, giving a reduction of 90.6\% (Fig.~S17). Each comparison uses the same endpoints, lattice vectors, atom mapping, and interatomic potential. The lower maxima therefore result from a different sequence of structural rearrangements: the direct pathways transform the periodic structure cooperatively, whereas the interface-mediated pathways proceed through local reconstruction and subsequent phase-front propagation.

The M(100)$\rightarrow$T(100) comparison further shows how the available propagation length affects the pathway. In the long structure, which extends approximately 12~nm along the interface normal, the interface-mediated pathway reaches 136.03~meV per formula unit. The direct cooperative pathway in the corresponding short structure, approximately 3~nm along the same direction, reaches 192.10~meV per formula unit, giving a reduction of 29.2\% (Fig.~S18). The two structures retain the same phase orientations and in-plane lattice vectors but differ in their length normal to the interface. The long structure accommodates sequential phase-front motion, while the shorter periodic structure transforms more cooperatively. This comparison therefore captures the change in transformation mechanism enabled by the additional propagation length.

Together, the NEB profiles connect the observed interface behavior with the directional accessibility of the corresponding structural pathways. The strong M-directed asymmetry of M(001)/T(100), the weaker asymmetry of the pinned M(100)/T(100) boundary, the near balance between PO and AO, and the AO-directed T/AO pathway are consistent with the low-temperature phase distributions. The fixed-lattice NEB profiles characterize potential-energy pathways and are used here to compare directional barriers and transformation mechanisms, while the temperature-dependent phase populations are obtained independently from the MD simulations.

\section*{Discussion}

Phase selection at coherent HfO$_2$ boundaries reflects both the relative bulk free energies and the structural compatibility of each crystallographic matching. Within the same parent-phase pair, changing the orientation can preserve coexistence, promote conversion toward one parent, or produce a different local phase mixture. This result is consistent with previous studies showing that surfaces, domain walls, coherent boundaries, and strain can alter the relative stability of competing hafnia polymorphs \cite{Materlik2015,Park2017Surface,Ding2020DomainWall,Falkowski2021Interphase,Xu2024Strain,Kelley2023}. The present 45-interface survey further resolves how these effects evolve from 300 to 1800~K. A scale estimate based on published coherent M/T boundary energies also suggests that interfacial contributions may remain relevant at nanometer dimensions (Supplementary Information).

The temperature-dependent phase maps separate the interfaces into two broad response classes. All non-M interfaces become T-dominant at 1800~K, whereas every M-containing interface retains a monoclinic majority and a substantial T population. The persistence of M is consistent with the coupled monoclinic shear, asymmetric Hf--O coordination, and oxygen-sublattice rearrangement required for M$\rightarrow$T conversion. These distortions help preserve a connected monoclinic framework, as illustrated by M(100)/T(100), which remains mixed over 3~ns from 300 to 1800~K. The non-M interfaces respond more strongly to the organization of polar and antipolar displacements. PO can promote correlated polar order in an adjoining T region, while the response of AO depends on orientation and termination. In particular, the reconstruction associated with AO(010) is consistent with a locally polar termination initiating PO-like order. Separating PO and AO in the local phase analysis was therefore necessary to distinguish retention of antiparallel order from conversion toward a polar structure.

The pathway calculations connect these phase distributions to the available structural rearrangements. The directional barriers around the relaxed interface states are consistent with the low-temperature tendencies toward M, balanced PO/AO coexistence, or AO retention. Same-lattice comparisons show that the pathways through relaxed interface states reduce the PO(001)$\rightarrow$AO(001) and T(001)$\rightarrow$AO(001) maxima by 72.3\% and 90.6\%, respectively. For M(100)$\rightarrow$T(100), increasing the length along the interface normal from approximately 3 to 12~nm produces a 29.2\% reduction and allows sequential phase-front motion instead of a more cooperative transformation. These results suggest that small periodic structures may suppress pathways requiring progressive reconstruction over an extended region. Earlier calculations showed that the T-to-M barrier depends strongly on whether monoclinic shear is permitted \cite{Xu2021}; the present comparison indicates that the available propagation length also affects the resulting pathway. A related role for extended T/O configurations was reported by Lee \textit{et al.} in field-driven HZO, where transient T regions mediate the formation of oppositely polarized orthorhombic domains \cite{Lee2026FieldInduced}. Although the material and applied conditions differ, both studies indicate that intermediate phase regions can participate directly in structural transformation.

The present conclusions apply to coherent, stoichiometric, and defect-free HfO$_2$ interfaces with prescribed orientations. The 3~ns simulations establish persistence under the simulated conditions, while equilibrium interface free energies and experimental lifetimes require separate assessment. Similarly, the fixed-lattice NEB profiles compare potential-energy pathways rather than finite-temperature rates. Future studies incorporating oxygen vacancies, dopants, electric fields, electrodes, and interacting grain boundaries could determine how these factors modify interface stability and phase-front mobility. Within the present scope, crystallographic orientation, interface termination, and available propagation length emerge as distinct variables for retaining metastable phase combinations, pinning phase boundaries, or promoting directed phase conversion in HfO$_2$-based materials.

\section*{Methods}

\begin{figure*}[htbp]
	\centering
	\includegraphics[width=\textwidth]{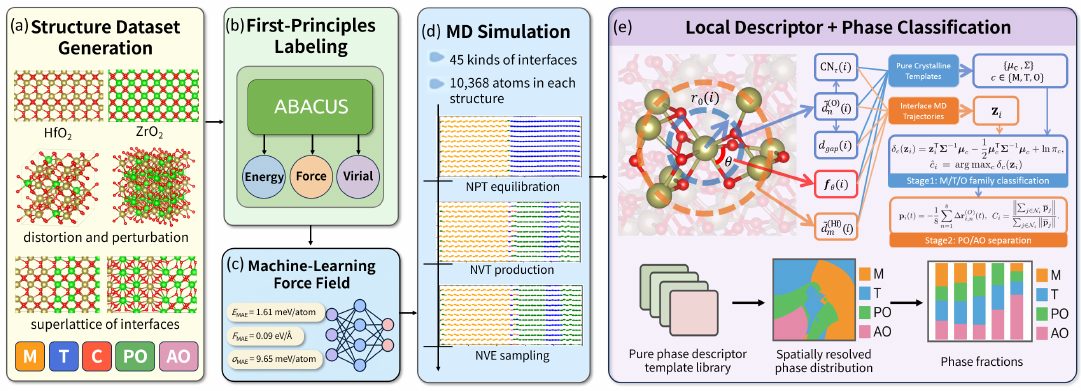}
	\caption{\textbf{Computational workflow.} (a) Structural dataset construction covering HfO$_2$, ZrO$_2$, and Hf$_x$Zr$_{1-x}$O$_2$ polymorphic configurations and interfacial superlattices with strain, distortion, and random perturbations. (b) First-principles labeling of energies, forces, and virials. (c) Machine-learning force field training. (d) Large-scale MD simulations of 45 two-phase interface models with 10,368 atoms each. (e) Local phase classification: a 5~ps average of the 21-component Hf-centered descriptor is assigned by equal-prior shrinkage linear discriminant analysis to M, T, or the O family; O sites are then separated into PO and AO using the 15~ps local polar coherence of Hf off-centering vectors.}
	\label{fig:FrameWork}
\end{figure*}

The overall workflow is illustrated in Fig.~\ref{fig:FrameWork}, proceeding from dataset construction and first-principles labeling to potential training, molecular dynamics of interface models, and local phase analysis.
The atomic structures in this work were visualized using VESTA \cite{MommaVESTA2011}.

\subsection*{Dataset Preparation and Machine-Learning Force Field}

First-principles calculations were performed using the ABACUS code with the PBESol functional \cite{Li2016,Chen2010,Perdew2008}. To build a transferable force field for polymorphic hafnia, the dataset includes crystalline and interfacial superlattice configurations of HfO$_2$, ZrO$_2$, and Hf$_x$Zr$_{1-x}$O$_2$, covering the M, T, PO, AO, and C phases, together with strained, distorted, and randomly perturbed structures. ZrO$_2$ was included because it shares the same major polymorphic families as HfO$_2$ and provides additional local environments relevant to phase competition and interfacial reconstruction. The final dataset contains 60,192 configurations. Detailed first-principles settings are provided in the Supplementary Information.

The DP force field was trained using the DeePMD-kit framework \cite{Wang2018,Zhang2020,Zeng2023} by simultaneously fitting DFT energies, atomic forces, and virials. Transferability to interfaces was evaluated by two complementary tests: applying the frozen force field to every frame of 4~ps NPT AIMD trajectories for M(100)/T(100) and T(100)/PO(001) cells with $\sim$10~\AA lattice parameter at 300, 900, and 1800~K, and DFT relabeling of 600 configurations from independent 1~ns DPMD trajectories. The comparable energy and force errors between these two sets confirm that the force field reproduces both AIMD-sampled and self-generated interfacial structures. Details on sampling, convergence, and parity plots are given in the Supplementary Information.

\subsection*{Quasi-Harmonic Approximation Free Energy Calculations}

Bulk Gibbs free energies of the M, T, PO, AO, and C phases were evaluated within the quasi-harmonic approximation. For each phase, static energies and phonon frequencies were calculated for uniformly scaled volumes around the equilibrium structure. The Helmholtz free energy, $F(V,T)=E_0(V)+F_{\mathrm{vib}}(V,T)$, was minimized with respect to volume at zero pressure to obtain $G(T)$. DFT calculations used ABACUS and Phonopy, and the same volume sampling and minimization were applied to the DP force field using phonoLAMMPS, LAMMPS, and Phonopy \cite{Li2016,Chen2010,CarrerasPhonoLAMMPS2020,Togo2010,phonopy-phono3py-JPSJ,LAMMPS}.

\subsection*{Two-Phase Interface Models and Atomistic Simulations}

Large-scale molecular dynamics simulations of two-phase HfO$_2$ interfaces were performed using the trained DP force field with LAMMPS \cite{LAMMPS}. The interface set comprises 45 coherent superlattice models built from M, T, PO, and AO polymorphs with interface normals along the [100], [010], and [001] directions. Each model contains 10,368 atoms. The two constituent slabs were combined into a common supercell by adopting the average in-plane lattice constants of the two parent phases, while retaining the out-of-plane lattice constants of the individual slabs along the stacking direction. These stoichiometric periodic cells deliberately isolate coherent HfO$_2$/HfO$_2$ boundary geometry, omitting electrodes, free surfaces, point defects, and composition gradients, thereby enabling a controlled comparison of boundary-mediated reconstruction.

The standard finite-temperature protocol consists of 100~ps NPT, 200~ps NVT, and 20~ps NVE sampling. The temperature series spans 300 to 1800~K in 300~K increments. Long-time records contain three consecutive 1~ns segments; ensemble sequences and restart checks are provided in the Supplementary Information.

\subsection*{Local Phase Classification and Polar Order Assignment}

A two-step classifier was employed to resolve the four HfO$_2$ polymorphs under finite-temperature broadening. First, a 21-component Hf-centered descriptor (combining normalized Hf--O distances, O--Hf--O angular statistics, and normalized Hf--Hf distances) was averaged over 5~ps for each persistent Hf site and assigned to M, T, or the orthorhombic family by equal-prior shrinkage linear discriminant analysis \cite{LedoitWolf2004}. No spatial averaging was applied.

Orthorhombic-family sites were subsequently separated into PO and AO. For each site, a Hf off-centering vector relative to the centroid of the eight nearest O neighbors was averaged over a rolling 15~ps interval, and its local $q=0$ coherence over the central site and its 80 nearest Hf neighbors was evaluated. Parallel polar order yields a large coherence (PO), while antiparallel cancellation yields a small coherence (AO). Temperature-specific thresholds were calibrated using only phase-pure, phase-preserving fixed-cell trajectories from perfect crystals; no interface trajectories were used for training or threshold selection. The final output contains exactly the M, T, PO, and AO classes. Independent validation at all six temperatures confirmed correct classification, and transfer to joined interfaces recovered the two parent slabs while preserving localized reconstruction at the boundary. Full descriptor definitions, calibration details, and validation results are provided in the Supplementary Information and Fig.~S9.

\subsection*{Fixed-Lattice Interface State NEB Calculations}

NEB calculations were performed for selected interface orientations using the DP force field. For each orientation, the relaxed interface state and its two pure phase endpoints were represented in the same periodic cell containing 1,152 atoms, with fixed lattice vectors, atom ordering, and endpoint coordinates. Bands were optimized with the improved-tangent ordinary NEB formulation \cite{HenkelmanJonsson2000} implemented in the Atomic Simulation Environment \cite{LarsenASE2017}. For the two M/T orientations, independently refined Dimer saddles \cite{HenkelmanJonsson1999} served as fixed shared endpoints of two NEB half-bands to preserve the intended I$\rightarrow$M or I$\rightarrow$T connectivity. For the PO/AO and T/AO interfaces, full interface-to-endpoint bands were optimized directly.
Direct PO(001)$\rightarrow$AO(001) and T(001)$\rightarrow$AO(001) controls used the same 1,152-atom endpoint pairs, fixed-lattice, and atom mapping as the corresponding branches through the interface. More details are provided in the Supplementary Information.

\section*{Data availability}

The data of this study are publicly available at \url{https://github.com/xdzhu/HfO2_interface_project}. The raw trajectories from the large-scale MD simulations are not included because of the large volume but are available from the corresponding author upon reasonable request.

\section*{Code availability}

The code used in this study is publicly available at \url{https://github.com/xdzhu/HfO2_interface_project}. ABACUS is an open-source DFT code distributed under the GPL 3.0 license and is available at \url{http://abacus.ustc.edu.cn}. DeePMD-kit is an open-source deep-learning potential package distributed under the GPL 3.0 license and is available at \url{https://github.com/deepmodeling/deepmd-kit}. LAMMPS is open-source molecular dynamics software distributed under the GPL 2.0 license and is available at \url{https://www.lammps.org}.

\section*{Acknowledgements}
This work was supported by the Anhui Provincial Science and Technology Breakthrough Project (Grant No. 202523o09050015), the Operating Funds of the Institute of Artificial Intelligence, Hefei Comprehensive National Science Center (Grant No. 22KT004), and the Innovation Program for Quantum Science and Technology (Grant No. 2021ZD0301200). Numerical computations were performed on the USTC HPC facilities and at the Hefei Advanced Computing Center.

\section*{Funding}
The Anhui Provincial Science and Technology Breakthrough Project (Grant No. 202523o09050015) is funded by the Department of Science and Technology of Anhui Province. The Operating Funds of the Institute of Artificial Intelligence, Hefei Comprehensive National Science Center (Grant No. 22KT004) are funded by the Institute of Artificial Intelligence, Hefei Comprehensive National Science Center. The Innovation Program for Quantum Science and Technology (Grant No. 2021ZD0301200) is funded by Hefei National Laboratory.

\section*{Author contributions}
X.Z. and L.H. conceived the project. X.Z. and J.L. performed the calculations and analyzed the results. X.Z. and L.H. wrote the manuscript. L.H. supervised the research. All authors commented on the manuscript.

\section*{Competing interests}

The authors declare no competing financial or non-financial interests.

\bibliography{HfO2_interface-v2}

\section*{Figure legends}

\noindent\textbf{Fig.~1 | Relative Gibbs free energies of bulk HfO$_2$ polymorphs.}
Relative Gibbs free energies, $\Delta G_{\alpha}(T)=G_{\alpha}(T)-G_{\mathrm{M}}(T)$, of the M, T, C, PO, and AO phases calculated within the quasi-harmonic approximation using DFT and the DP force field. Solid and dashed curves denote the DFT and DP force field results, respectively. Orange, blue, red, green, and pink denote the M, T, C, PO, and AO phases, respectively. The M phase defines the zero-energy reference at each temperature. Energies are reported in meV per formula unit (f.u.).

\noindent\textbf{Fig.~2 | Representative M/T interface structures after the 300~K finite-temperature protocol.}
For each interface, the upper and lower panels show projections onto the $ac$ and $bc$ planes, respectively. Phase-colored Hf sites are assigned using the M, T, PO, and AO local descriptor classification. (a) M(100)/T(100) retains two extended parent domains separated by a relatively sharp boundary. (b) M(001)/T(100) transforms predominantly into M, leaving a narrow T-like boundary. (c) M(100)/T(001) develops a reconstructed layer containing M, PO, AO, and residual T environments. (d) M(010)/T(001) reconstructs into a mixed stacking dominated by M and PO environments.

\noindent\textbf{Fig.~3 | Phase fractions of the two-phase interface models at representative temperatures.}
Phase fractions of 45 two-phase interface models at (a) 300~K, (b) 900~K, (c) 1500~K, and (d) 1800~K. The models are grouped by initial parent phase pair into the M/AO, M/PO, M/T, PO/AO, T/AO, and T/PO families. Colors indicate the fractions of Hf sites assigned to local M (orange), T (blue), PO (green), and AO (pink) environments by the 21-component local phase classifier.

\noindent\textbf{Fig.~4 | Nanosecond phase evolution of two representative interfaces.}
(a--c) M(100)/T(100) at (a) 300~K, (b) 900~K, and (c) 1800~K. (d--f) T(100)/PO(001) at (d) 300~K, (e) 900~K, and (f) 1800~K. Stacked areas show the M, T, PO, and AO fractions assigned by the 21-component local phase classifier over nonoverlapping 5~ps windows.

\noindent\textbf{Fig.~5 | Transformation pathways through representative relaxed interface states.}
Fixed-lattice energy profiles for (a) M(001)$\rightarrow$T(100), (b) M(100)$\rightarrow$T(100), (c) PO(001)$\rightarrow$AO(001), and (d) AO(001)$\rightarrow$T(001). Energies in each panel are reported relative to the corresponding interface state (I), and the annotations indicate the directional barrier for each pathway segment.

\noindent\textbf{Fig.~6 | Computational workflow.}
(a) Structural dataset construction covering HfO$_2$, ZrO$_2$, and Hf$_x$Zr$_{1-x}$O$_2$ polymorphic configurations and interfacial superlattices with strain, distortion, and random perturbations. (b) First-principles labeling of energies, forces, and virials. (c) Machine-learning force field training. (d) Large-scale MD simulations of 45 two-phase interface models with 10,368 atoms each. (e) Local phase classification: a 5~ps average of the 21-component Hf-centered descriptor is assigned by equal-prior shrinkage linear discriminant analysis to M, T, or the O family; O sites are then separated into PO and AO using the 15~ps local polar coherence of Hf off-centering vectors.

\end{document}